\documentclass[12pt,preprint]{aastex}
\usepackage{amsmath,amssymb,graphicx,aas_macros}

\newsavebox{\astrutbox}
\sbox{\astrutbox}{\rule[-5pt]{0pt}{20pt}}

\newcommand{\gapprox}{\lower.4ex\hbox{$\;\buildrel >\over{\scriptstyle\sim}\;$}}
\newcommand{\lapprox}{\lower.4ex\hbox{$\;\buildrel <\over{\scriptstyle\sim}\;$}}

\begin{document}

\title{MAGNETIC HELICITIES AND DYNAMO ACTION IN MAGNETO-ROTATIONALLY DRIVEN TURBULENCE} 
 
 \author{G. Bodo\thanks{E-mail:
bodo@oato.inaf.it}~, F. Cattaneo$^{2}$, A. Mignone$^{3}$ and P. Rossi$^{1}$\\
$^{1}$INAF/Osservatorio Astrofisico di Torino, Strada Osservatorio 20, 10025 Pino Torinese, Italy\\
$^{2}$Department of Astronomy and Astrophysics, The University of Chicago, 
              5640 S. Ellis avenue, Chicago IL 60637, USA\\
$^{3}$Dipartimento di Fisica, Universit\`a degli Studi di Torino, Via Pietro Giuria 1, 10125 Torino, Italy}

% \begin{document} 
 \date{Accepted ??. Received ??; in original form ??}
  
%%%%%%%%%%%%%%%%%%%%%%%%%%%%%%%%%%%%%%%%%%%%%%%%%%%%%%%%%%%%%%%%%%%%%%%%%%%% 
\begin{abstract} 
We examine the relationship between magnetic flux generation, taken as an indicator of large-scale dynamo action, and magnetic helicity, computed as an integral over the dynamo volume,  in a simple dynamo. 
We consider dynamo action driven by  Magneto-Rotational Turbulence  (MRT) within the shearing-box approximation. We consider magnetically open boundary conditions that allow a flux of helicity in or out of the computational domain. We circumvent the problem of the lack of gauge invariance in open domains by choosing a particular gauge -- the winding gauge -- that provides a natural interpretation in terms of average winding number of pairwise field lines. We use this gauge precisely to define and measure the helicity and helicity flux for several realizations of dynamo action.
We find in these cases, that the system as  a whole does not break reflectional symmetry and the total helicity remains small even in cases when substantial magnetic flux is generated. We find no particular connection between the generation of magnetic flux and the helicity or the helicity flux through the boundaries. We suggest that this result may be due to the essentially nonlinear nature of the dynamo processes in MRT.
\end{abstract} 
%%%%%%%%%%%%%%%%%%%%%%%%%%%%%%%%%%%%%%%%%%%%%%%%%%%%%%%%%%%%%%%%%%%%%%%%
%\begin{keywords}
%accretion disk , MHD, dynamos, turbulence 
 %\end{keywords}

\section{Introduction}
The magnetic helicity is a measure of the topological complexity of a magnetic field; in an ideal fluid bounded by a co-moving flux surface, it is a conserved quantity. This results makes precise the idea that in the absence of diffusion, magnetic field lines cannot reconnect and therefore, all their knots and linkages must be preserved \citep{moffatbook}. By contrast, dynamo action describes the sustained generation of magnetic fields by inductive processes within the bulk of an electrically conducting fluid; it often involves efficient reconnection or enhanced diffusivity, and more often than not, involves substantial changes in field topology.
In a plasma with high electrical conductivity these two processes must find some way to compromise. How this compromise is reached is the subject of considerable debate. 
If the magnetic diffusivity is small, or more precisely, if the magnetic Reynolds number $Rm$ is large, one would expect that the field topology be nearly preserved, or, alternatively,  that substantial topological changes might require timescales of the order of the (long) diffusion time. This is  incompatible with the requirement that a dynamo operating at high $Rm$ should be able to change the field topology on the (short) dynamical time-scale. The problem is particularly acute in the case of large-scale dynamo action. The latter describes the generation of substantial amounts of magnetic flux. One can imagine that if the flux bundles making up the magnetic field are at all linked, increasing the flux in the bundles by dynamo action will lead to a corresponding increase in magnetic helicity in blatant violation of the general idea that in a highly conducting fluid the magnetic helicity should be (nearly) conserved.  This has lead to a sustained theoretical effort to find ways to reconcile helicity conservation on the one hand, and efficient flux generation on the other  \citep{Kulsrud92, Cattaneo92, Diamond94, Cattaneo96, Field02, Blackman02, Shukurov06, Sur07}. 

One popular choice is based on the observation that if the region in which the dynamo operates is not enclosed by flux surfaces then helicity need not be conserved. If the magnetic field ``pokes through'' some of the boundaries, under the right conditions, there could be a non-zero flux of helicity in or out of the dynamo region that could alleviate the constraint posed by helicity conservation and allow efficient flux generation \citep{Blackman00, VC01, Subramanian04, Shukurov06, Ebrahimi14}. At some basic level the idea is that the magnetic field could keep itself reasonably disentangled by expelling all the unwanted tangles though the boundaries. The particularly attractive feature of this idea is that the process of expulsion does not require fast reconnection or turbulent diffusivity, in principle,  it could operate even at infinite $Rm$. It should be noted that if this general idea is correct one would expect the presence of open boundaries to be a necessary condition for flux generation, and some evidence of a connection between large-scale dynamo action and helicity fluxes through the boundaries. In theory, such a connection could be sought in numerical models of large-scale dynamos with open boundaries in an effort better to understand the relationship between dynamo action and helicity expulsion. One problem, however, immediately arises that makes a straightforward application of these considerations problematic. In a domain with open magnetic boundaries the helicity is not gauge invariant, and consequently nor are the fluxes.  Different choices of gauge could lead to widely different results thereby undermining  the whole analysis.
Fortunately, there appear to be at least two natural ways to circumvent this problem. One is to replace the helicity with some other quantity that is related to the helicity, in other words it is still a measure of topological complexity,  that is gauge invariant in an open field configuration. This avenue was explored by \citet{Berger84} and led to the introduction of the {\it relative helicity}. One would hope that even though the relative helicity is not quite the helicity it may still be informative about dynamo action. The other is to choose a particular gauge and stick to it. This approach is very simple and very straightforward, but it requires some justification of why a particular gauge is selected, and more importantly some understanding of what is the geometric or topological interpretation of the helicity in that specific gauge \citep{Prior14}. In particular one would want to choose a gauge that is relevant to dynamo processes. 

We address some of these issues in the present paper. Our approach is straightforward. We choose for our analysis the results of two numerical studies of dynamo action driven by the Magneto-Rotational-Instability (MRI) in shearing-box simulations (\citealp{Bodo14,Bodo15}, \citealp[for a general discussion on MRI see e.g.][]{Balbus03}). One of the reasons for choosing this particular setup is that with appropriate boundary conditions, the geometry of shearing-boxes is such that the radial and azimuthal magnetic fluxes are not conserved. Thus large-scale dynamo action can be easily demonstrated, and indeed it is,  by monitoring the generation of these fluxes. We then choose a specific gauge that leads to a natural topological interpretation for the resulting helicity,   compute the helicity  and the helicity fluxes in this gauge and compare them with dynamo activity.  We would like to note that other quantities that involve  the magnetic vector potential and that may have topological interpretations have been proposed as  relevant to dynamo action in various geometries. A notable example is the Vishniac and Cho flux \citep{VC01} that may be important in the accretion disk geometry. In this paper, however, we choose to focus  on the magnetic helicity alone. We believe this is an appropriate starting point since it is the magnetic helicity (properly defined as an integral quantity) that is both an ideal invariant and carries topological information about the magnetic field.
Finally we close the paper with a discussion of our findings.

\section{The Model}
We will perform the analysis of magnetic helicity behavior for two sets of stratified shearing box simulations of MRI driven turbulence with zero net vertical magnetic flux, already presented elsewhere \citep{Bodo14, Bodo15}. The two sets differ essentially for the equation of state: In the first set we employed an isothermal equation of state \citep{Bodo14}, while in the second set we used  the perfect gas equation of state with a finite thermal diffusivity \citep{Bodo12, Bodo13, Bodo15}. The magnetic and mechanical boundary conditions are the same for both: periodic in the azimuthal direction $y$, shear periodic in the radial direction $x$, while in the vertical direction we assume that the boundaries are in hydrostatic balance, impenetrable and stress free, giving 
\begin{equation}
v_z = 0, \qquad \frac{\partial v_x}{\partial z} = \frac{\partial v_y}{\partial z} = 0,
\end{equation}
and also that the magnetic field is purely vertical, giving
\begin{equation}
\frac{\partial B_z}{\partial z} = 0, \qquad B_x = B_y = 0.
\end{equation}
We should note that these conditions allow a net flux of magnetic helicity through the boundaries.  It is instructive to compare these two cases because they give rise to different kinds of dynamo action. In the perfect gas cases, there are situations in which the dynamo is of the large-scale type capable of generating substantial amounts of toroidal flux, whereas in the isothermal cases the dynamo always appears to be small-scale. Although it is clear that the differences in dynamo activity are associated with the development of vigorous convection in some of the perfect gas cases, why the presence of convection in MRI driven turbulence should enable the generation of large-scale fields is not currently understood. 

Both sets of simulations were performed with the PLUTO code \citep{PLUTO}.  As is often the case in many astrophysical simulations there is no  explicit viscosity or magnetic diffusivity, this is done to extract the largest possible dynamic range for the given computational resources. However, it does not mean that the evolution of the system is ideal, amongst other things that would preclude the possibility of dynamo action, rather it implies that dissipative processes are controlled by the resolution. Thus, for the simulations to be presented here,  the characteristic size of dissipative structures is of the order of a few gridpoints. This property notwithstanding, numerical dissipations acts very much like physical dissipation from the point of view of mediating reconnection and allowing changes in magnetic field topology. Furthermore, since PLUTO is written in conservative form for the total energy, whatever kinetic or magnetic energy is lost to numerical dissipation it is correctly accounted for in terms of a corresponding increase in internal energy. 

Finally, we note here that in the stratified shearing-box geometry, gravity is anti-symmetric about the mid-plane. As a result, quantities like the horizontally averaged {\it kinetic}  helicity density is likewise antisymmetric about the mid-plane, so in a sense, the regions above and below the mid-plane are individually not reflectionally symmetric, even though the layer as a whole is. However, in the cases we consider here, the normalized horizontally averaged  kinetic helicity density is very small, smaller than half of a percent throughout most of the interior, increasing to approximately 2 percent in thin boundary layers near the horizontal boundaries. So  the systems we consider are almost reflectionally symmetric. 

\subsection{The isothermal case}

\begin{figure}
  \centering
   \includegraphics[width=12cm]{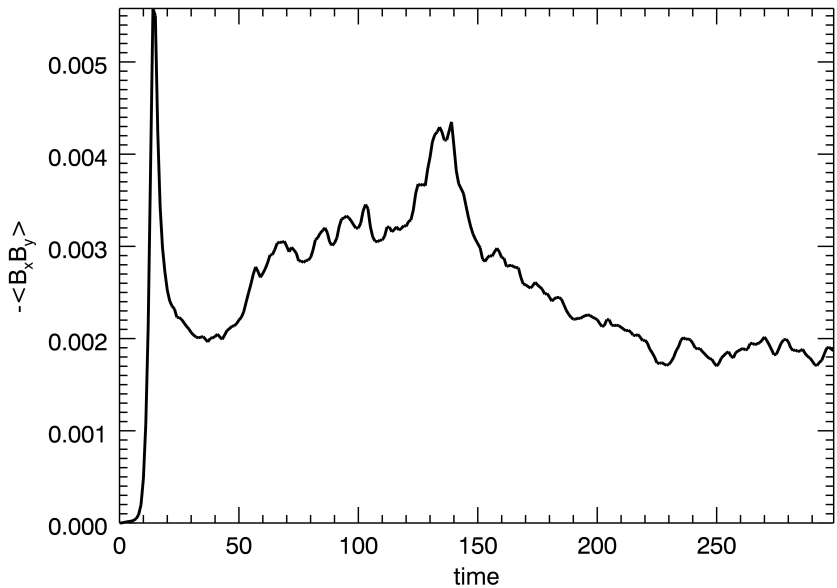} % requires the graphicx package
 \caption{ Isothermal case. Volume average of the $(x-y)$ component of the  Maxwell stress as a function of time. In these simulations, the Maxwell stress gives the dominant contribution to the outward transport of angular momentum and gives a measure of the efficiency of the dynamo action. The time  is measured in units of  $\Omega^{-1}$ and the stresses in units of $\rho_0 H^2 \Omega^2$. In these units a rotation corresponds to $2 \pi$ time-units.
 }
   \label{fig:stresses_iso}
\end{figure}

The computational domain covers the region
 \begin{equation}
-0.5 H < x  < 0.5 H, \qquad 0 < y  <\pi H, \qquad -3 H < z < 3 H,
\end{equation}
where $H$ is the scale height.  The grid size is $128 \times 384 \times 768$  with 128 grid-points per scale height.  As mentioned above, there is no net vertical magnetic flux threading the box. The equilibrium density distribution is given by
\begin{equation}\label{eq:density}
\rho = \rho_0 \exp(-z^2/H^2).
\end{equation}
The solutions are characterized by cyclic patterns of mean azimuthal field propagating away from the equatorial plane and becoming more evident at high altitudes, while the equatorial region is dominated by small scale fluctuations. This behavior has been observed in all isothermal simulations \citep[see e.g.][]{Davis10, Shi10, Gressel10, Flaig10, Guan11}  and is  best described in term of two coupled dynamos operating in different regions: a small-scale turbulent dynamo operating in the mid-plane region, where most of the mass is concentrated, and a mean-field type dynamo that operates in the tenuous overlying regions and generates magnetic structures in the form of upward propagating dynamo waves \citep{Gressel10}. Our data set covers an interval of time of $300 \;  \Omega^{-1}$, where $\Omega$ is the shearing box rotational frequency.  In  Fig. \ref{fig:stresses_iso} we plot the resulting volume averaged Maxwell stresses (in units of $\rho_0 H^2 \Omega^2$) as a function of time.  

\subsection{The perfect gas case}

\begin{figure}
   \centering
   \includegraphics[width=12cm]{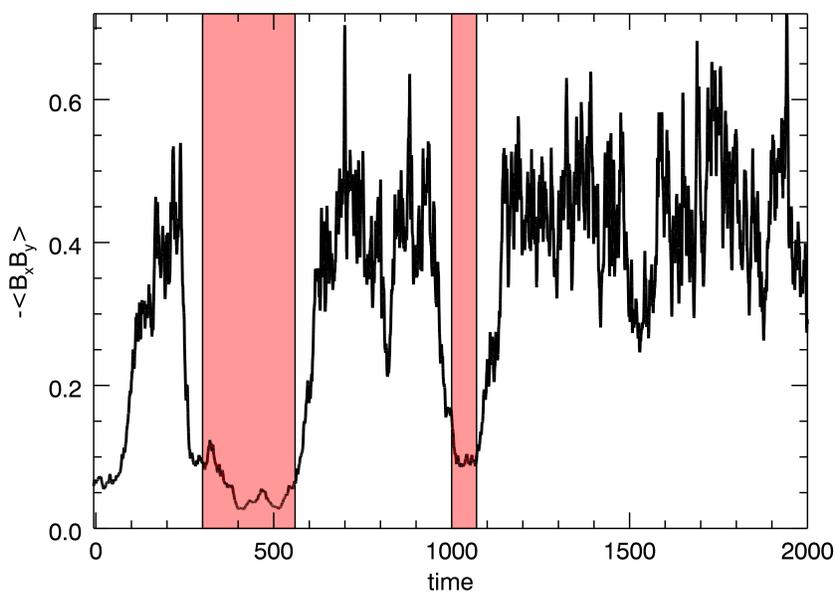} % requires the graphicx package
   \caption{Perfect-gas case. Volume average of the $(x-y) $ component of the  Maxwell stress as a function of time. 
The red shaded regions mark the time intervals in which the system is in the antisymmetric state. 
The time  is measured in units of  $\Omega^{-1}$ and the stresses in units of $\rho_0 H^2 \Omega^2$. In these units a rotation corresponds to $2 \pi$ time-units.
}
   \label{fig:stresses}
\end{figure}

In the second series of simulations we adopted the perfect gas equation of state, and also introduced thermal conduction so that the heat generated by dissipation of the MRI turbulence can be carried to the top and bottom  boundaries, where it is radiated away. The simulations are initialized with an isothermal equilibrium configuration, with a density distribution given by Eq. \ref{eq:density}, and are then  evolved until they reach a stationary configuration in which all the heat dissipated by turbulence is radiated away at the top and bottom boundaries. The computational domain covers the region  
 \begin{equation}
-8 H < x  < 8 H, \qquad -6 H < y <  6 H, \qquad -2 H < z < 2 H,
\end{equation}
where $H$ is the scale height of the initial isothermal atmosphere. The final quasi-steady state is completely different from the initial one and bears no memory of it, however we keep a reference to the scale height of the initial isothermal configuration in order to make possible a comparison between these simulations and the isothermal ones. The computational domain is covered by a grid of $512 \times 384 \times 128$ grid-points. In this case, because of the introduction of thermal conduction, we need an additional boundary condition, in addition to those described above,  for specifying either the temperature or the heat flux at the boundaries. Our choice has been to put the heat flux equal to the black body energy flux at the surface temperature, namely at $z = \pm 2 H$ we impose
\begin{equation}
\frac{\gamma}{\gamma-1} \kappa \rho {\cal R} \frac{d T}{d z} = \mp \sigma T^4
\end{equation}
where $T$ is the temperature, $\kappa$ is the thermal diffusivity, $\cal R$ is the gas constant and $\sigma$ is the Stefan-Boltzmann constant. We found two distinctive regimes: conductive and convective, corresponding respectively to large and small values of the thermal diffusivity. In the conductive regime, the heat generated by dissipation is transported by thermal conduction, the temperature and density profiles are not very different from  the isothermal cases and also dynamo action has similar  properties. At low thermal diffusivities the layer becomes unstable to convective motions that  carry all the heat, the density profiles become flat and we observe an extremely effective large-scale dynamo capable of generating substantial amounts of toroidal flux and, correspondingly, much larger values of the Maxwell stresses.  This behavior has been also confirmed by \citet{Gressel13, Hirose14}  and \citet{Hirose15}.  Two broad classes of dynamo solutions can be identified: the ones in which the toroidal flux is symmetric about the mid-plane and the others in which it is anti-symmetric. The symmetric solutions have higher Maxwell stresses.  The system appears to swap between these states randomly but, typically, spends more time in the symmetric, higher-efficiency states. The data set covers a time interval of $2000 \; \Omega^{-1}$.  In Fig. \ref{fig:stresses} we show the behavior of  the volume averaged Maxwell stresses (in units of $\rho_0 H^2 \Omega^2$) as a function of time;  we have marked in red the time intervals in which we observe the antisymmetric solution.

\section{Magnetic Helicities}
The primary objective of the present work is to explore the relationship between  magnetic helicity  and large-scale dynamo activity.
Before this task can be undertaken we need to provide ourselves with a workable definition of the magnetic helicity in a domain with open magnetic boundaries. Let ${\bf B}$ and ${\bf A}$ be the magnetic field intensity and the associated vector potential satisfying
\begin{equation}
{\bf B}= \nabla \times {\bf A} \quad {\rm with} \quad \nabla \cdot {\bf B} =0.
\label{eq:solenoidal}
\end{equation}
Clearly the above expression does not define ${\bf A}$ uniquely but only up to the addition of the gradient of an arbitrary scalar function. Thus 
the gauge transformation 
\begin{equation}
{\bf A} \rightarrow {\bf A} + \nabla \Psi
\label{eq:gauge}
\end{equation}
leaves ${\bf B}$ unchanged. In an ideal fluid the evolution of ${\bf B}$ is governed by the induction equation
\begin{equation}
\partial_t {\bf B} =\nabla \times ({\bf u}\times {\bf B}), 
\label{eq:induction}
\end{equation}
where {\bf u} is the fluid velocity. Using (\ref{eq:solenoidal}) we can uncurl the induction equation to obtain a corresponding evolution equation for the vector potential, namely
\begin{equation}
\partial_t {\bf A} =({\bf u}\times {\bf B})+ \nabla \Phi,
\label{eq:uncurled}
\end{equation}
where $\Phi$ is an arbitrary function related to the the choice of gauge. The magnetic helicity is defined by 
\begin{equation}
H=\int_{V} dV ({\bf A} \cdot {\bf B}),
\label{eq:helicity}
\end{equation}
where the integral is over the volume containing the magnetofluid. The evolution equation for $H$ can readily be obtained from (\ref{eq:induction}) and (\ref{eq:uncurled}),
\begin{equation}
\frac{dH}{dt} = -\int_{S} (F + F_a) dS \quad {\rm with} \quad F =({\bf B} \cdot {\bf n})({\bf A} \cdot {\bf u} +\Phi) \quad {\rm and }  \quad
F_a=({\bf A} \cdot {\bf B} ) ({\bf u} \cdot {\bf n}),
\label{eq:hel-flux}
\end{equation}
where $S$ is the surface bounding $V$ and ${\bf n}$ is the outward normal. Clearly, if $S$ is a comoving, or impenetrable boundary $F_a$ vanishes, furthemore, 
if $S$ is a flux surface, i.e. $({\bf B} \cdot {\bf n})=0$ on $S$ then $F=0$ and the helicity is conserved and gauge invariant. On the other hand, if $({\bf B} \cdot {\bf n}) \neq 0$ on some subset of $S$ there will be a corresponding contribution to $F$ and the helicity will, in general, not be a conserved quantity. Notice that in this case, the flux of helicity through the boundaries defined by $F$, and therefore the helicity itself depend on the choice of gauge. This is a well known problem associated with domains with open magnetic boundaries. As we mentioned in the introduction there are at least two natural ways to proceed. One is to give up the helicity altogether and to replace it with some other quantity that, still has a useful topological interpretation, and is gauge invariant. This approach was pursued by \citet{Berger84} who noted that the difference between the helicities associated with two distinct magnetic fields could be made  gauge independent. This opens the possibility to define in a gauge invariant way the helicity in an open field domain relative to some reference field. Although in principle the reference field need only satisfy some appropriate boundary conditions, a popular choice is a potential field in which case the resulting quantity is called the relative helicity. 
Because a potential field is current free, a possible topological interpretation for the relative helicity is that it  is the helicity associated with the current distribution in the fluid. For a domain bounded by plane surfaces, the relative helicity is defined by
\begin{equation}
H^R=\int_{V}({\bf A} + {\bf A_p })({\bf B} - {\bf B_p}) dV
\label{eq:relative}
\end{equation}
where ${\bf B}_p$ is the potential field with the same normal component as ${\bf B}$ on the boundaries, and ${\bf A}_p$ is the associated vector potential. 

Another possibility is to pick a particular gage. Given our geometry, a good choice is given by the winding gauge, discussed by \citet{Prior14}  and defined by the requirement that the vector potential satisfies
\begin{equation}
\nabla_H \cdot {\bf A}^w = 0, \quad {\rm where} \quad \nabla_H \equiv (\partial_x, \partial_y, 0).
\label{eq:winding}
\end{equation}
It is possible to show that in cylindrical domains, i.e. domains in which the vertical boundaries are flux surfaces while the horizontal ones are open, the helicity computed with the winding gauge gives the average pairwise winding of the field lines \citep{Prior14}. This gives a particularly intuitive picture of the field topology. To see how the helicity and helicity flux can be computed in the winding gauge we begin by noting that in general
\begin{equation}
{\bf J} = \nabla \times {\bf B} = \nabla \times (\nabla \times {\bf A})=-\nabla^2 {\bf A} + \nabla(\nabla \cdot {\bf A} ).
\label{eq:def-A}
\end{equation}
which upon use of (\ref{eq:winding}) and restricting to our geometry, gives
\begin{equation}
{\bf J} = -\nabla^2 {\bf A}^w + \nabla (\partial_z A_z^w).
\label{eq:vec-winding}
\end{equation}
The $z$ component of equation (\ref{eq:vec-winding})  reads 
\begin{equation}
J_z=-\nabla_H^2 A_z^w
\label{eq:A3}
\end{equation}
which can be solved to find $A_z^w$ up to an additive function of $z$ , $f(z)$, say. $A_z^w$ can then be substituted back into (\ref{eq:vec-winding}) to obtain the horizontal components of ${\bf A}^w$. We note that, unlike $A_z^w$,  $A_x^w$ and $A_y^w$ are uniquely defined. 
The scalar potential $\Phi^w$ in the winding gauge can likewise be computed  by taking the horizontal divergence of (\ref{eq:uncurled}), which gives
\begin{equation}
\nabla_H^2 \Phi^w =-\nabla_H \cdot ({\bf u} \times {\bf B}).
\label{eq:phi}
\end{equation}
The last equation defines $\Phi^w$ up to an additive function of $z$, $g(z)$, say. We now show that for our geometry, the winding helicity $H^w$ and the associated boundary fluxes are independent of the choice of $f$ and $g$. We consider a transformation of the form

\begin{equation}
{\bf A}^w \rightarrow {\bf A}^w + {\bf e}_z f(z), \quad \Phi^w\rightarrow \Phi^w +g(z),
\label{eq:gauge-1}
\end{equation}
its effect on $H^w$ is to give rise to an extra term of the form
\begin{equation}
\int_{V} dV f(z) B_z= \int dz f(z) \int dx dy B_z = 0,
\label{eq:inv_1}
\end{equation}
since there is no net vertical flux. Similarly from (\ref{eq:hel-flux}) the contributions to the helicity flux have the form
\begin{equation}
f(z) \int_{S} dx dy B_z u_z  + g(z) \int_S dx dy B_z =0,
\label{eq:in_2}
\end{equation}
where the first integral vanishes because $u_z$ vanishes on the horizontal boundaries, and the second vanishes because, as before, there is no net vertical flux.

%%%----------------------- This is the piece that deals with the cases with net horizontal flux  --------------------------
Inspection of equation (\ref{eq:A3}) shows that some care is required when the vertical component of the current vanishes identically. This is, for instance, the case for magnetic fields of the form 
\begin{equation}
{\bf B} = (b_1(z), b_2(z), 0).
\label{eq:hor-filed}
\end{equation}
From which it follows that the $z$ component of the vector potential in the winding gauge can be written as 
\begin{equation}
A_z=C_1xb_2(z)+C_2yb_1(z),
\label{eq:Az}
\end{equation}
for any constants $C_1$ and $C_2$. Substituting (\ref{eq:Az}) into (\ref{eq:vec-winding}) immediately gives the horizontal components of the vector potential in the form
\begin{equation}
A_x=(C_1+1)G_2(z), \qquad A_y=(C_2-1)G_1(z),
\label{eq:AxAy}
\end{equation}
where $G_1'(z)=b_1(z)$ and $G_2'(z)=b_2(z)$. Clearly the resulting vector potential satisfies the winding gauge requirement (\ref{eq:winding}) and is compatible with the prescribed magnetic field for any choice of $C_1$ and $C_2$. This ambiguity can easily be removed by considering the specific case of a horizontal field whose amplitude is constant and whose direction rotates uniformly with $z$. This can be written as
\begin{equation}
{\bf B}(z)=(\sin z, \cos z, 0).
\label{eq;unif-rot}
\end{equation}
It is reasonable to demand that the helicity density for such a field should be unity which corresponds to the specific choice $C_1=C_2=0$, which is what we adopt here. 
Finally, we note that adding a constant (vector) to the vector potential again satisfies all the requirements of the winding gauge but leads to changes in the helicity density. Again the ambiguity can be removed by requiring that uniform horizontal magnetic fields have vanishing helicity density which gives that the (vector) constant must itself be zero. 
%%%----------------------- This is the piece that deals with the cases with net horizontal flux  --------------------------

The previous considerations apply to an ideal fluid. If non ideal effects are included, like a finite magnetic diffusivity, $\eta$, say, the helicity is no longer conserved even in a magnetically closed system. The helicity flux must be modified and a new production term appears.
The evolution equation for the helicity then becomes
\begin{equation}
\frac{dH}{dt}=-\int_{S} F_T dS + \int_{V}  W dV,  
\label{eq:hel-evol-diss}
\end{equation}
where
\begin{equation}
F_T = F + \eta ~ {\bf n} \cdot ( {\bf A} \times {\bf J} ) , \quad {\rm and} \quad W=-2\eta ~ ({\bf B} \cdot {\bf J}).
\label{eq:hel-terms-diss}
\end{equation}
In the numerical simulations being considered here,  $\eta$ is not explicitly known since the dissipation is entirely numerical. It can, however be estimated for any simulation by noting that in a steady state the energy production rate, $Q$, must balance the luminosity, $L$. The latter can be computed directly from the simulation from the expression
\begin{equation}
L= \int_{S} \sigma T^4 dS ,
\label{eq:blak-body}
\end{equation}
where the integrals are computed at the top and bottom boundaries and $\sigma$ is Stefan-Boltzmann constant. The former can be approximated by the expression
\begin{equation}
Q= \eta \int_{V} dV {\bf J}^2,
\label{eq:heating}
\end{equation}
where we have assumed that the viscous heating is small compared to Ohmic heating. This is motivated by noting that the numerical Prandtl number for our codes is close to unity, and that in our solutions the Maxwell stresses always substantially exceed the Reynolds stresses. As we shall see presently, even though this is only a working assumption, it gives reasonable estimates. 

Having defined the helicity in two possible ways, relative and winding, it is natural to ask which, if any, is more naturally related to dynamo processes. Since there is no completely convincing argument that one choice is far better than the other, the sensible thing to do  is to try them both. However  the helicity densities associated with the relative helicity or with the winding helicity are practically identical for our set of data. Thus from here on we discuss all relevant quantities in the winding gauge.

\section{Results and Analysis}
We begin our analysis by specifying the units we adopt to describe the results. In order to be able to compare the isothermal and perfect-gas results we use the same units in both cases. Unfortunately there is no single set of units that is entirely natural to both systems, nevertheless, and as a reasonable compromise we adopt $1/\Omega$ as the unit of  time, the isothermal scale height $H$ as the unit of  length, and the  mid-plane density $\rho_0$ as the unit of  density.  In the perfect-gas case $H$ and $\rho_0$ are computed relative to an isothermal atmosphere in hydrostatic balance with the same total mass and vertical extent. 

\begin{figure}
   \centering
   \includegraphics[width=12cm]{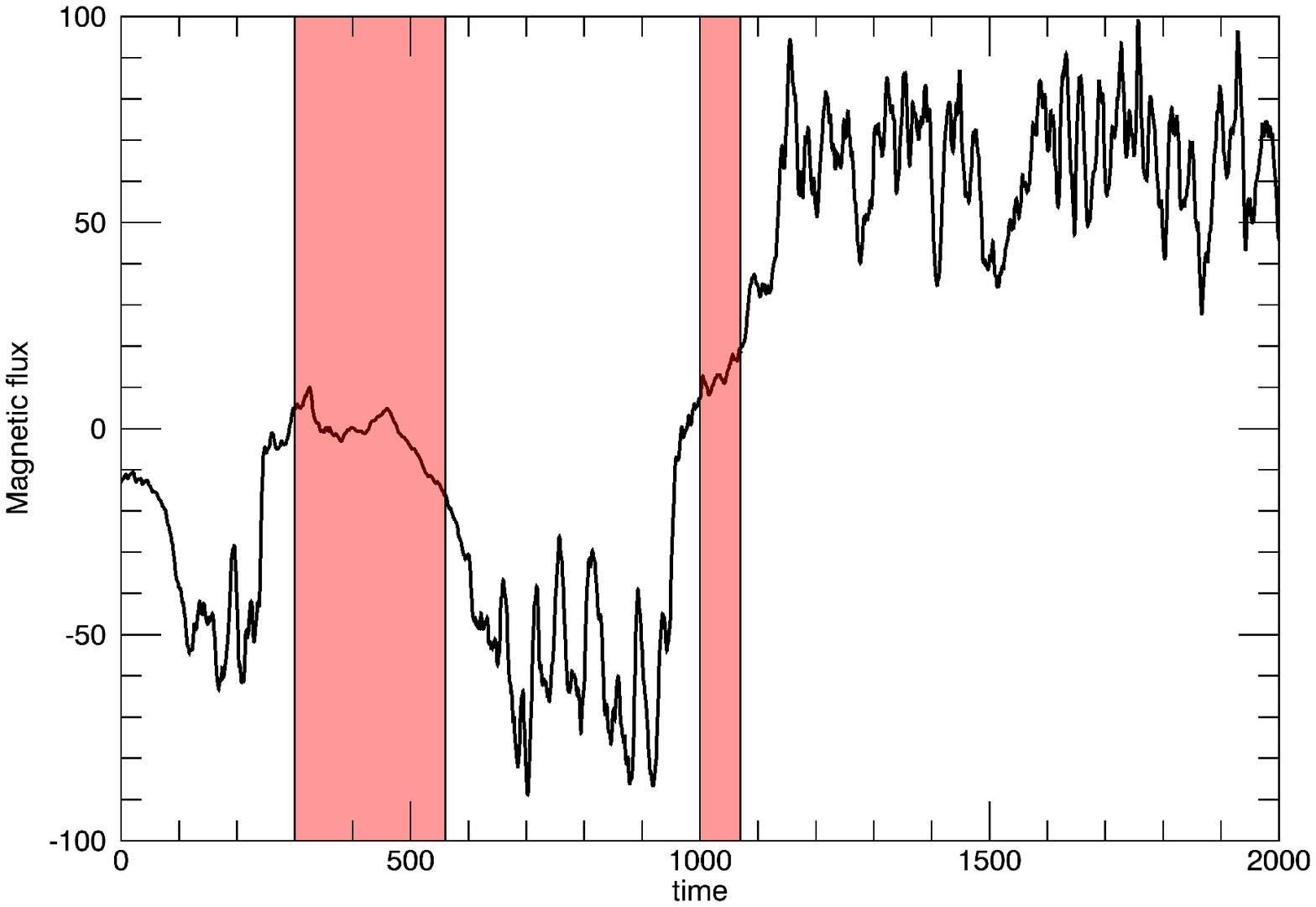} % requires the graphicx package
   \includegraphics[width=12cm]{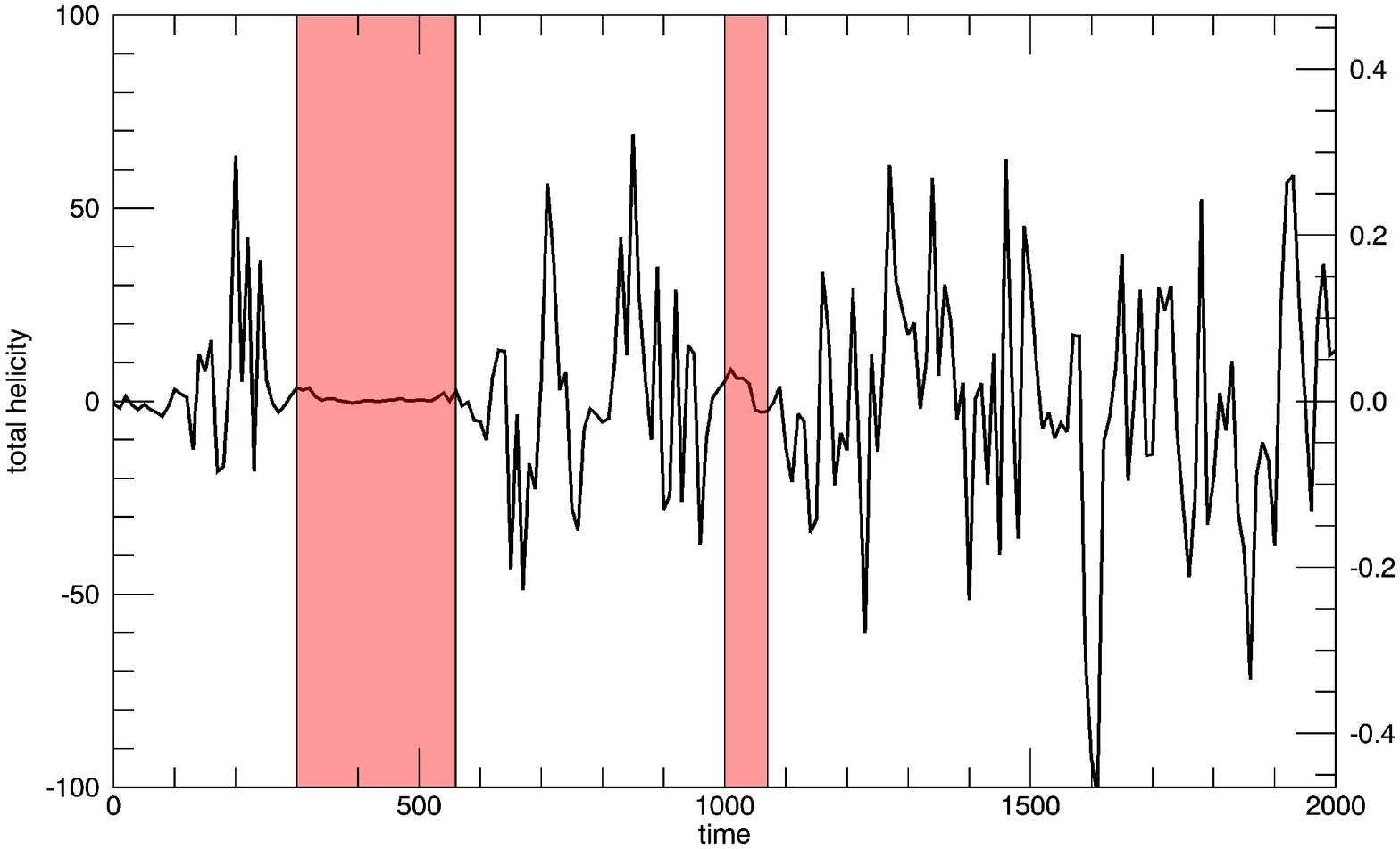} % requires the graphicx package
   \caption{Perfect-gas case. The upper panel shows the time history of the flux of the toroidal ($y$) component of the magnetic field. The lower panel shows the time history of  the helicity ($\int_V {\bf A}^w \cdot {\bf B}   dV$) for the corresponding time period. The scale on the right hand side gives the average twist between pairs of field lines. The regions shaded in red correspond to the anti-symmetric states. }
   \label{fig:tothel_conv}
\end{figure}

We consider the perfect-gas case first. All quantities are computed over the same time interval as in Figure~\ref{fig:stresses}. Figure~\ref{fig:tothel_conv} shows toroidal flux (upper panel) and  total helicity (lower panel) as functions of time. In order to get a feeling for the values of the helicity we recall that the helicity in the winding gauge measures the average twist between pairs of field lines in units of the rms square flux. Accordingly, on the right of the figure (lower panel) we have included a scale in these units. 

 \begin{figure}
   \centering
   \includegraphics[width=12cm]{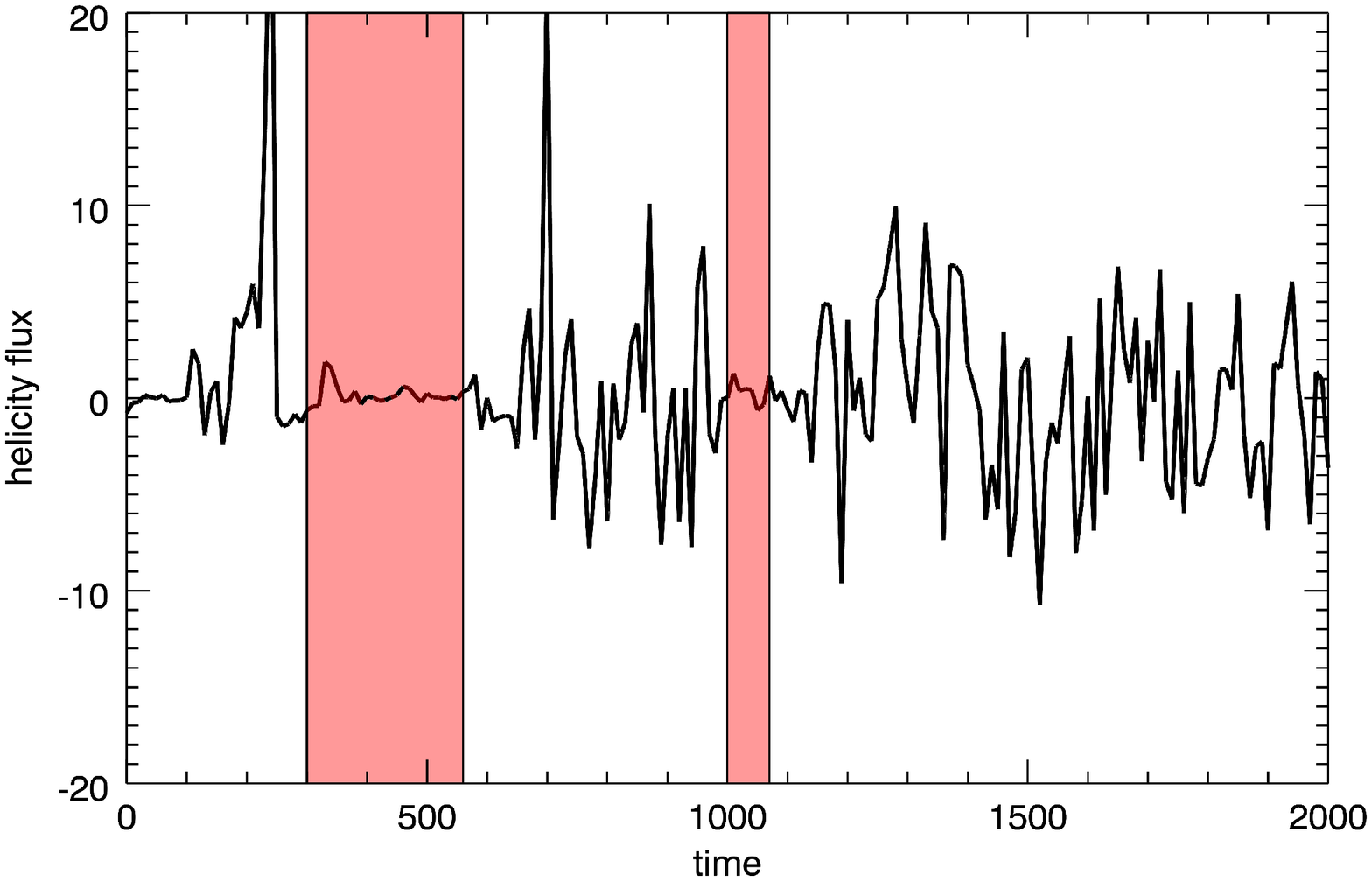} % requires the graphicx package
   \includegraphics[width=12cm]{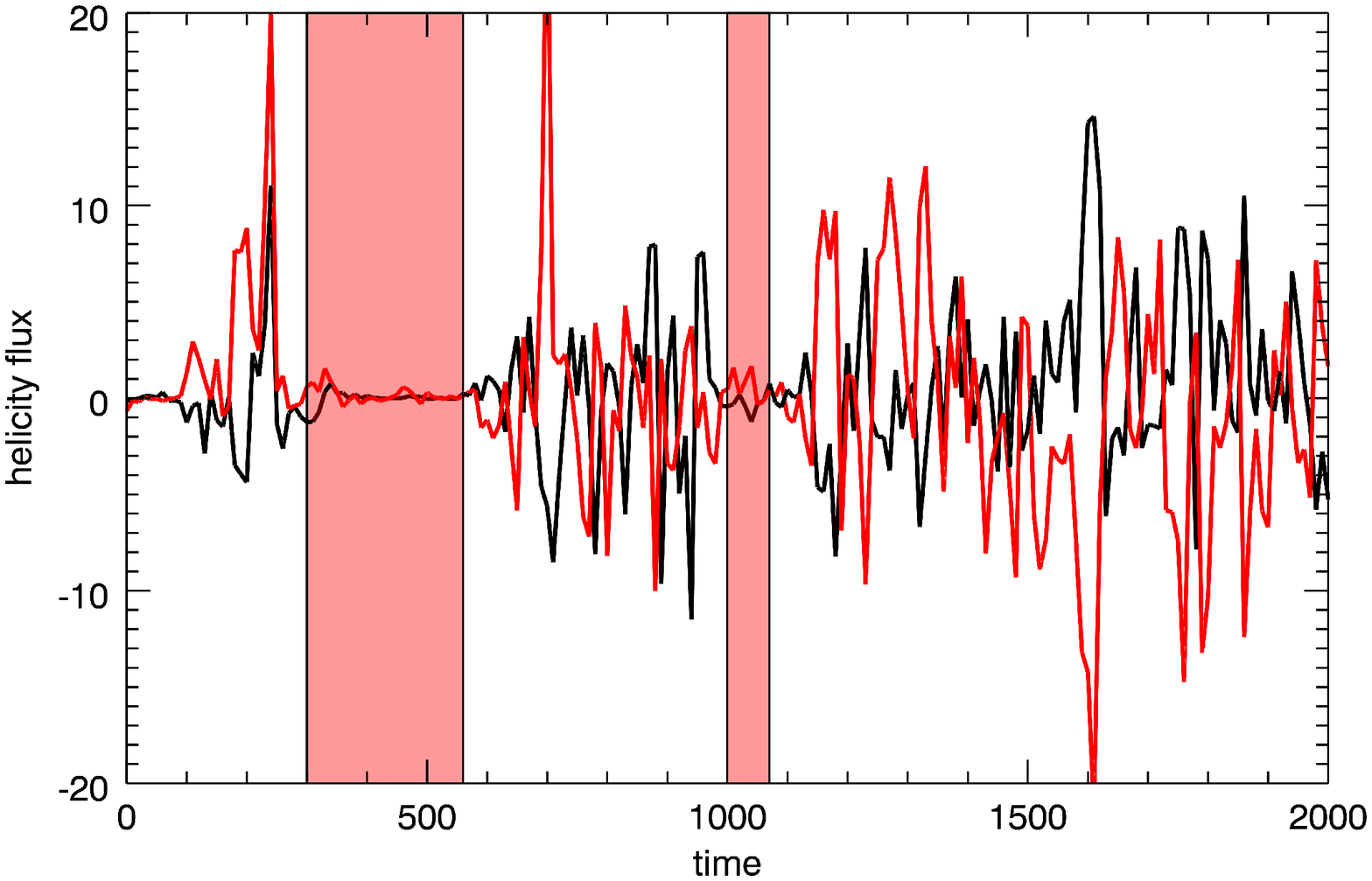} % requires the graphicx package
   
   \caption{Perfect-gas case. Time history of the helicity fluxes though the horizontal boundaries. 
The upper  panel shows the time history of the total helicity flux through the boundaries ${\mathcal F(t)}$. In the lower panel the contributions from the ideal part of the helicity flux (black curve) and from the  non-ideal part  (red curve) are shown separately. The regions shaded in red correspond to the anti-symmetric states. }
   \label{fig:helflux_conv}
\end{figure}

Some comments are in order. At a typical time, the average amount of twist is modest; less than unity actually. Assuming that the magnetic field is concentrated in bundles, and as we shall see presently this is not  a bad assumption, the small average twist arises not because the individual bundles are untwisted, rather because they are twisted with either sign roughly equally. Also, the total helicity changes sign with time and there is no evidence of symmetry breaking with a corresponding establishment of a long-lasting helical state of either sign. Occasionally, during the anti-symmetric phases the helicity drops to very small values. Interestingly the toroidal flux remains one signed for much longer periods than the helicity. Thus, if we use the toroidal flux as a measure of large-scale dynamo action, it seems to be uncorrelated, or at any rate unconstrained by the total helicity. For instance, after $t \approx 1100$ the toroidal flux remains substantial and positive while the helicity changes sign over ten times during the same period. Of course, one should entertain the possibility that the helicity is small and uncorrelated to dynamo action because most of it is efficiently expelled through the boundaries. This being the case, there should be some correlation between the time history of the toroidal flux and the helicity flux in or out of the domain. This possibility can easily be eliminated by inspection of the upper panel in figure~\ref{fig:helflux_conv}, which shows the corresponding time history of the total helicity flux (i.e. the difference of the fluxes though the two horizontal surfaces), defined by
\begin{equation}
{\mathcal F(t)} = \int_S F_T (t) dS.
\label{def:flux}
\end{equation}
Figure~\ref{fig:helflux_conv} also shows the separate contributions to the total flux arising from the ideal and non-ideal parts (lower panel). 
Very much like the helicity itself, the helicity flux shows rapid fluctuations and no particular preference for either sign. Also, since the area under a typical peak is comparable with a typical value of the helicity, we conclude that the helicity flowing in and out of the volume is comparable to the amount of helicity inside the volume. Again one could argue that, since what is being shown in figure~\ref{fig:helflux_conv} is the {\it total} helicity flux, the fluxes though the individual boundaries could be much larger but with a small difference. Actually that is not the case, and indeed, the individual fluxes are comparable to the total. 
The other contribution to the  helicity is the helicity production rate, related to non-ideal effects and  given by last term in (\ref{eq:hel-evol-diss}), namely
\begin{equation}
S_H (t) = \int_{V}  W dV.
%S_H (t) = \int_{V}  -2\eta  ({\bf B} \cdot {\bf J}) dV.
\label{eq:production}
\end{equation}

 \begin{figure}
   \centering
   \includegraphics[width=12cm]{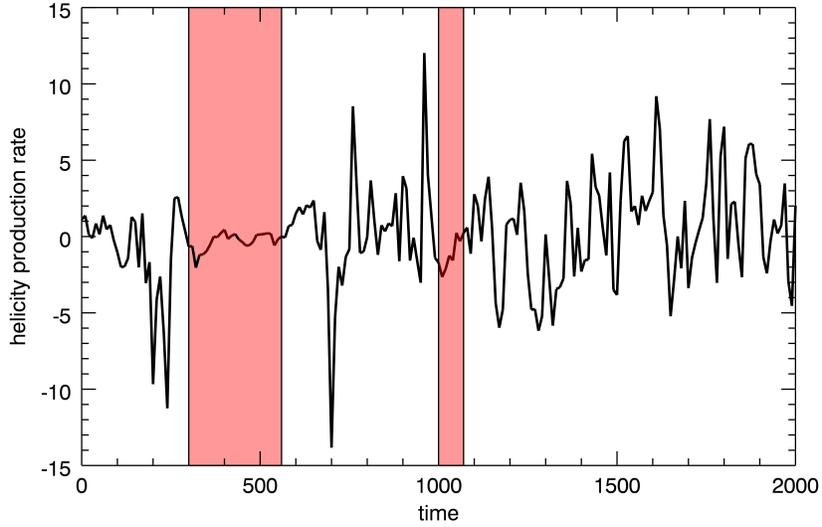} % requires the graphicx package
   \caption{Perfect-gas case. Time history of the helicity production rate computed from (\ref{eq:production}).  The regions shaded in red correspond to the anti-symmetric states.}
   \label{fig:helprod}
\end{figure}

\begin{figure}
   \centering
   \includegraphics[width=12cm]{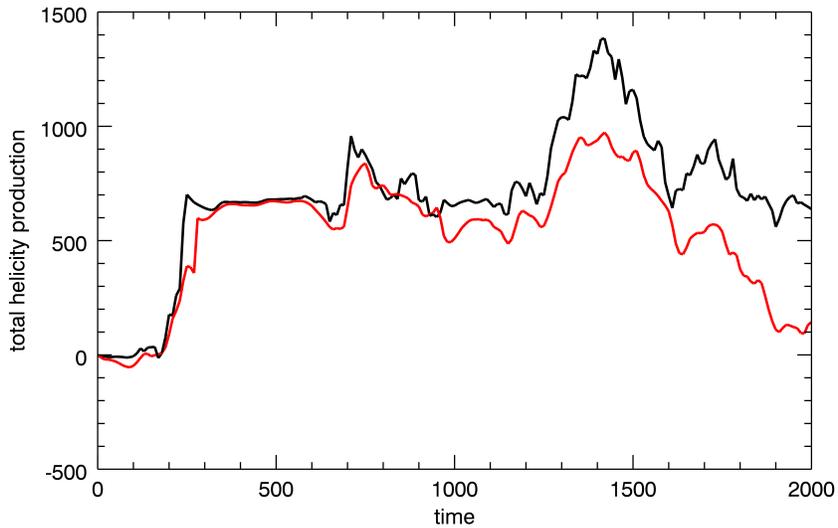} % requires the graphicx package
   \caption{Perfect-gas case. Time history of the magnetic helicity  produced  up to time $t$ by non-ideal effects. The red curve is computed by evaluating to the {\it right hand side} of expression (\ref{eq:thel}), the black line by evaluating the {\it left hand side}. Most likely, the discrepancies are due to numerical errors and limited time resolution.}
   \label{fig:helprodint}
\end{figure}

This measures how efficiently the turbulence itself can generate or destroy helicity. Figure~\ref{fig:helprod} shows the time history of $S_H(t)$  where we have used an effective diffusivity $\eta$ computed by equating the time averages of (\ref{eq:blak-body}) and (\ref{eq:heating}). Once again we note that the $S_H(t)$ fluctuates rapidly and has no particular sign. Furthermore, much as in the case of ${\cal F}(t)$, the helicity produced during a typical fluctuation is comparable to the total helicity in the domain. We should emphasize that this quantity is only a rough estimate, based on the assumptions that the unknown numerical dissipation can be approximated by  regular diffusion with an effective diffusivity estimated by the method above. 
To check the validity of this assumption we note that the time integral of (\ref{eq:hel-evol-diss}) gives 
\begin{equation}
H(t) + \int_0^t {\cal F}(t') dt' = \int_0^t S_H(t') dt'.
\label{eq:thel}
\end{equation}
Thus the helicity produced within the volume can also be estimated as the difference between the helicity present in the volume and that expelled (or injected) through the boundaries. Again we note that the above expression is only exact analytically, and not numerically. There are two main sources of possible errors: the first is that the code is not written in conservative form for the helicity, the second is that we only sample the results every unit of time, thus the (numerical) time integral is carried out on a coarsely sampled signal. The first error is controlled by the resolution the second by the cadence of the data collection. For our simulations the former is much, much smaller that the latter. 
These issues notwithstanding, the time histories of the helicity production computed from the two different methods are shown in figure~\ref{fig:helprodint}. 
Given the uncertainties in estimating the values of $\eta$, it is quite remarkable that the two curves track each other as well as they do.
It is clear that the curves slowly drift apart, but over a remarkably long time--of the order of several hundred orbits.   

%Clearly there are both periods during which the two curves more or less track each other as well as episodes in which a substantial %disagreement arises. We note in particular the event near $t=1,600$ in which the red curve reports that a considerable amount of helicity has %been generated (or destroyed) with no corresponding signal from the black curve. Our belief is that these discrepancies arise most likely from %the limited time resolution. 

\begin{figure}
   \centering
   \includegraphics[width=8cm]{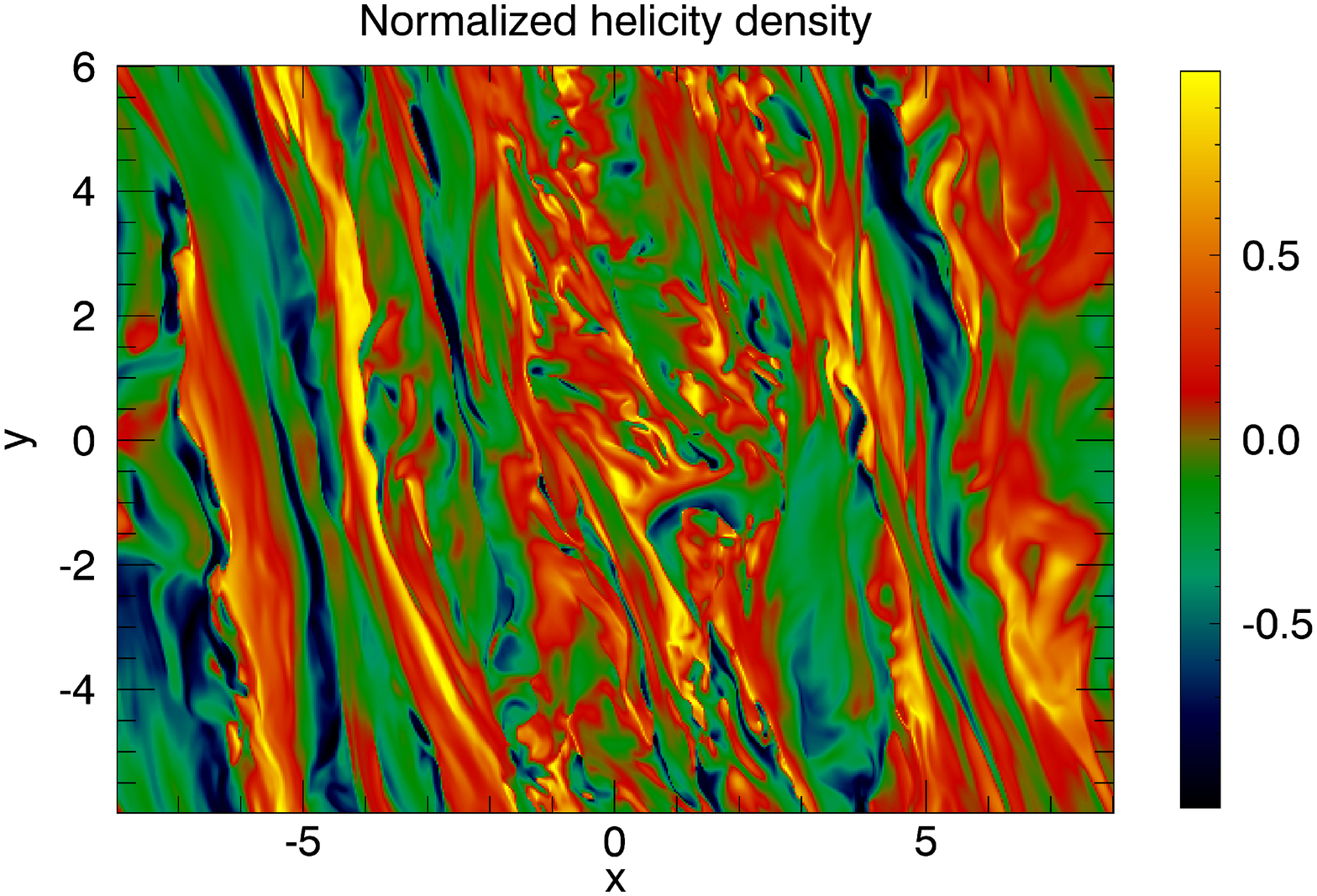} % requires the graphicx package
   \includegraphics[width=8cm]{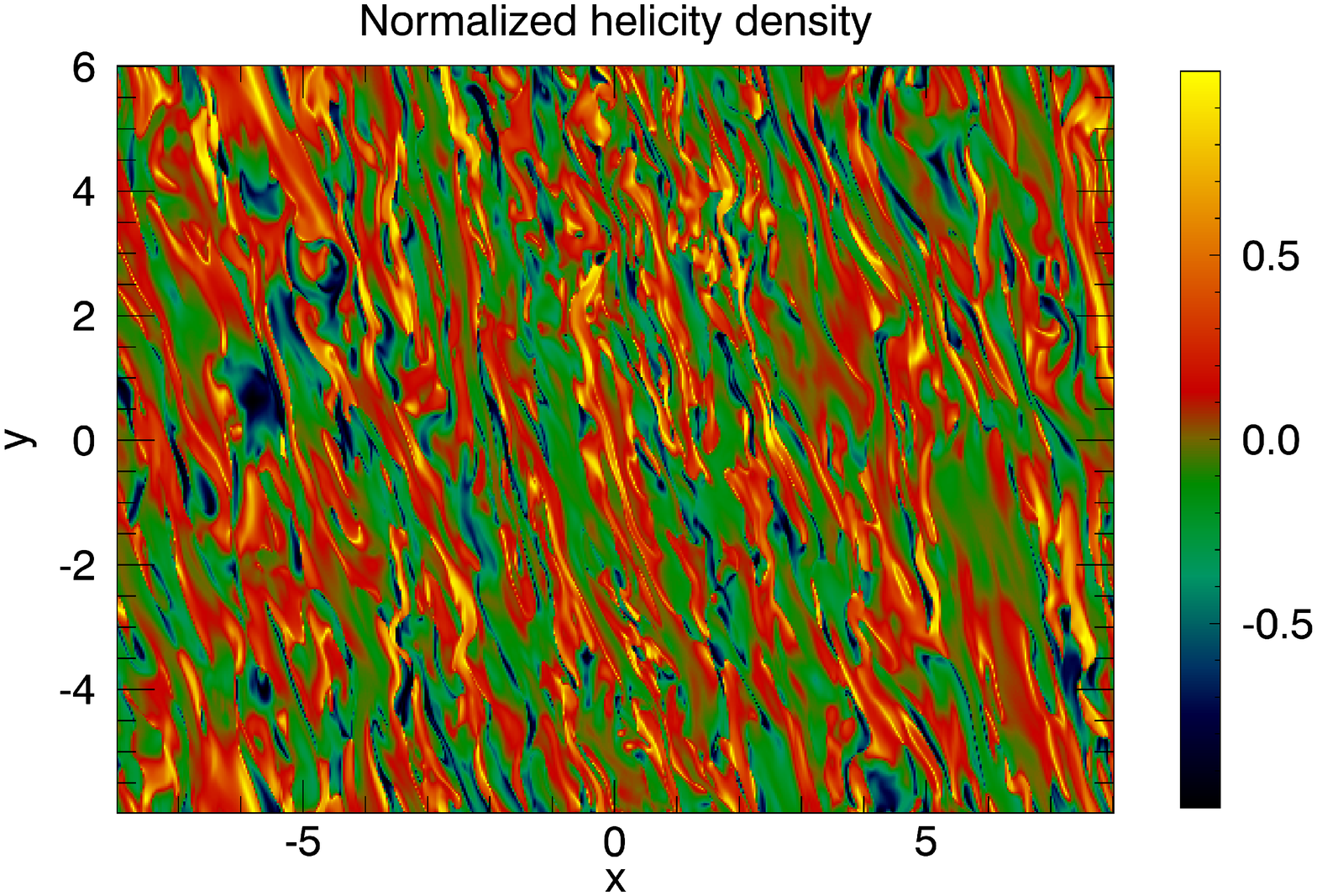} % requires the graphicx package
      \includegraphics[width=8cm]{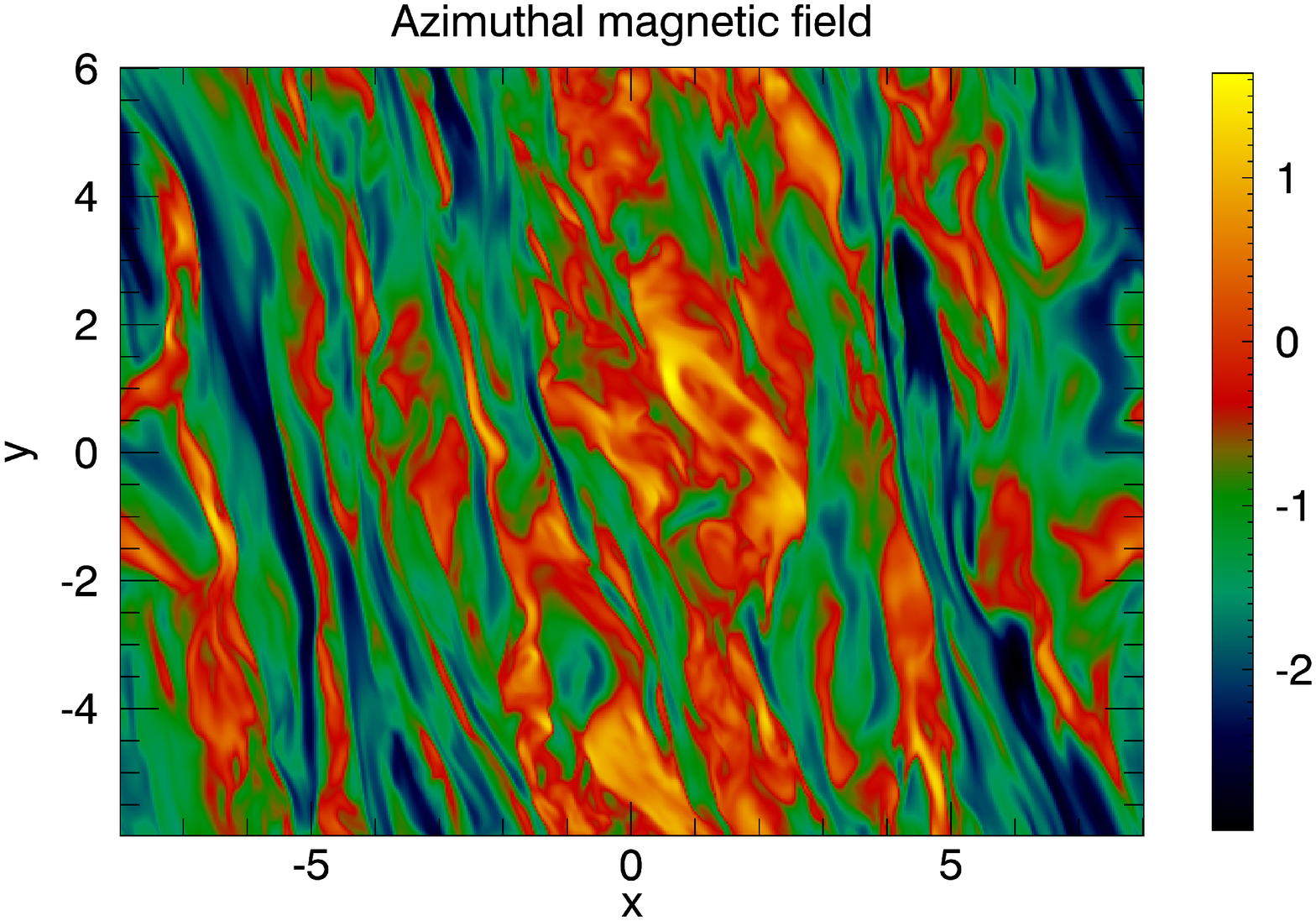} % requires the graphicx package
   \includegraphics[width=8cm]{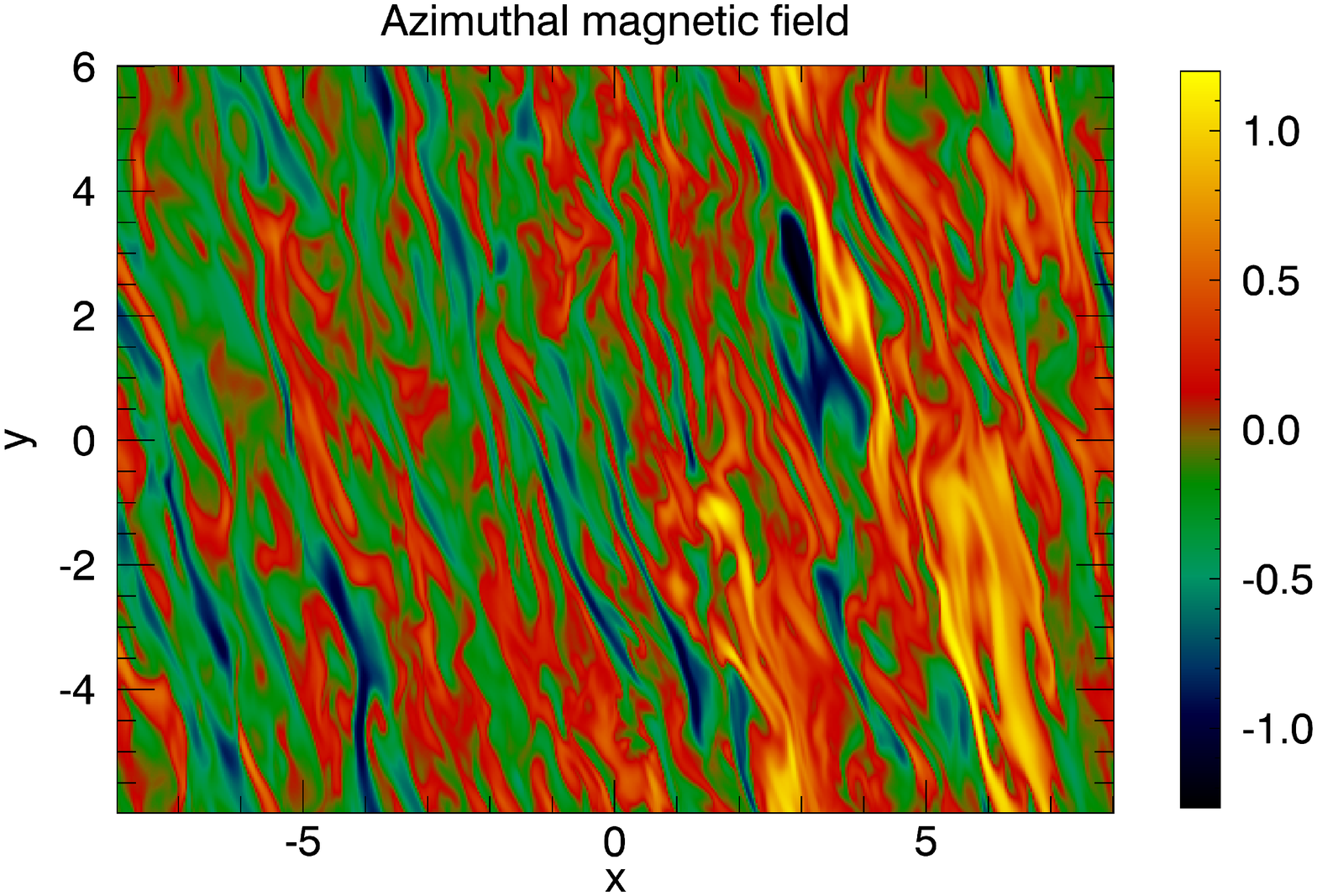} % requires the graphicx package

   \caption{Perfect-gas case. The two top panels show the spatial distribution of the normalized helicity density $( {\bf A} \cdot {\bf B} \ |{\bf A}| | {\bf B}|)$ in a horizontal  plane at $z=0$ at two different times.  The two bottom panels show the corresponding distributions of  the azimuthal magnetic field component. The left panels corresponds to a representative time in the symmetric state while the right panels corresponds to a representative time in the anti-symmetric state.  
}
   \label{fig:rhimage}
\end{figure}

\begin{figure}
   \centering
   \includegraphics[width=12cm]{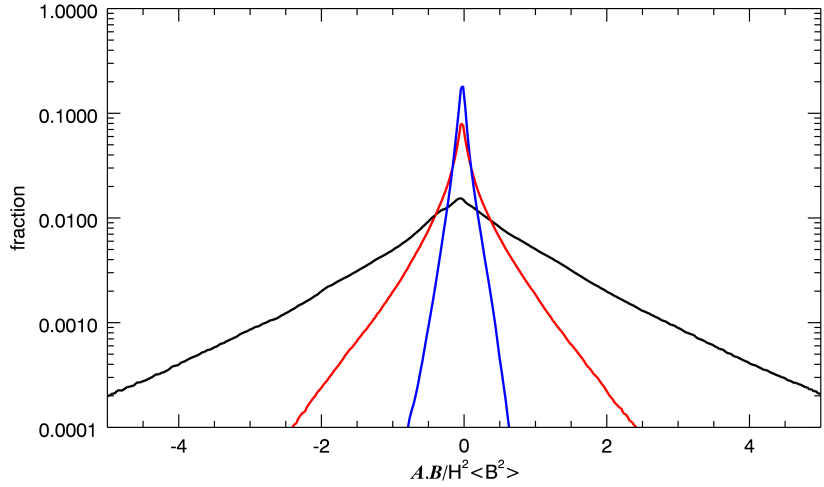} % requires the graphicx package
   \caption{Probability distribution functions of the helicity density normalized by $H^2 \langle {\bf B}^2 \rangle$. This corresponds to the average pairwise twist per unit length. Three cases are shown: perfect-gas, symmetric state--black curve; perfect-gas anti-symmetric state--red curve; isothermal case--blue curve. Each curve is computed at a representative time.}
   \label{fig:pdf}
\end{figure}

It is instructive to examine how the contributions to the total helicity are distributed spatially. To this end, figure~\ref{fig:rhimage} shows, in the two top panels, a  plot of the normalized helicity density in the mid-plane at two times during which the system is in the symmetric (left panel) and anti-symmetric state (right panel). In the same figure, in the two bottom panels,  we show also the corresponding distributions of the azimuthal magnetic field component, a comparison between the top and bottom panels show no apparent relationship between the two distributions.  Inspection of the top panels indicates that the helicity density is highly inhomogeneous and is concentrated in oblique sheet-like structures with both right and left handed twist. This left-right symmetry is further evidenced in figure~\ref{fig:pdf}  which shows the probability distribution function (pdf) of the helicity density  over the entire volume normalized by $< {\bf B}^2 >$ at the corresponding times. The black curve corresponds to the symmetric state, the red curve to the anti-symmetric state, and we shall return to the blue curve presently. Clearly the modest value of the total helicity arises mostly from the high degree of symmetry of the pdf and not by the fact that the pdf is narrowly distributed around zero. We can make this more precise by noting that for the symmetric case the rms value is approximately 25 times larger than the average value. If we apply this value of 25 to a typical fluctuation in figure~\ref{fig:tothel_conv} we conclude that, if all the flux bundles had the same sign, the average pairwise twist would be close to 2.5 which is also an estimate of the typical twist in a bundle. If we repeat a similar argument for the antisymmetric case, the average value is close to 12 times smaller than in the symmetric case, while the ratio of rms to average is 3.4 times larger which corresponds to a typical twist per bundle of 0.7 or approximately 3.5 times smaller than in the symmetric case. Inspection of figure~\ref{fig:rhimage} shows that the most striking difference between the symmetric and anti-symmetric cases is the the smaller size of the typical structure in the latter. This size difference can be used to argue that, in a simple-minded way, the anti-symmetric case can be thought of as two symmetric cases stuck on top of each other and squished back into the same original volume. In the horizontal, that would correspond to 4 copies of the symmetric layer next to each other and shrunk so as to cover the same original area. 

We note here, that the high degree of symmetry of the pdf's can be used to argue against the possibility that regions of positive and negative twist have different filling factors. One could conceive of a situation in which the positive twist, say, is gentle but occupies a large fraction of the volume, while the negative twist is high but concentrated over a smaller region so that the overall twist is small. This would manifest itself in a skewed distribution with nearly zero mean--which is definitely not the case here. 
We also note that the high degree of symmetry of the pdf's is preserved even when the distributions are computed separately for the regions above and below the mid-plane. This rules out possibility that the average twist is significant but distributed anti-symmetrically about the mid-plane. 

\begin{figure}
   \centering
   \includegraphics[width=12cm]{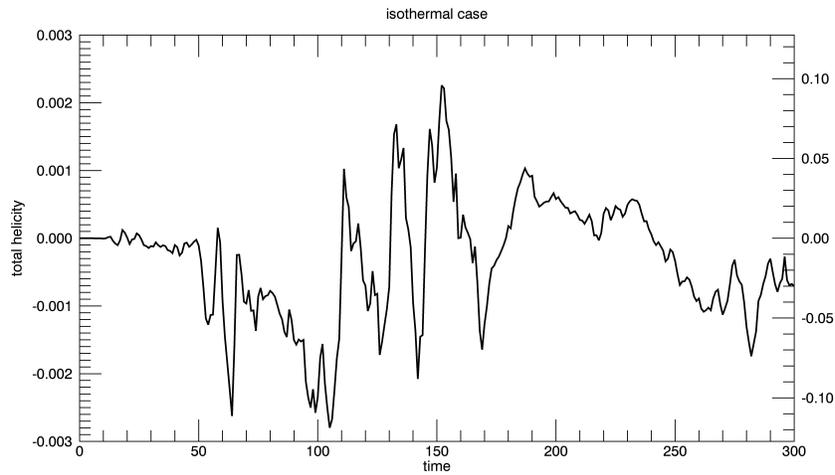} % requires the graphicx package
   \caption{Isothermal case. Time history of the helicity. }
   \label{fig:tothel_iso}
\end{figure}

\begin{figure}
   \centering
   \includegraphics[width=12cm]{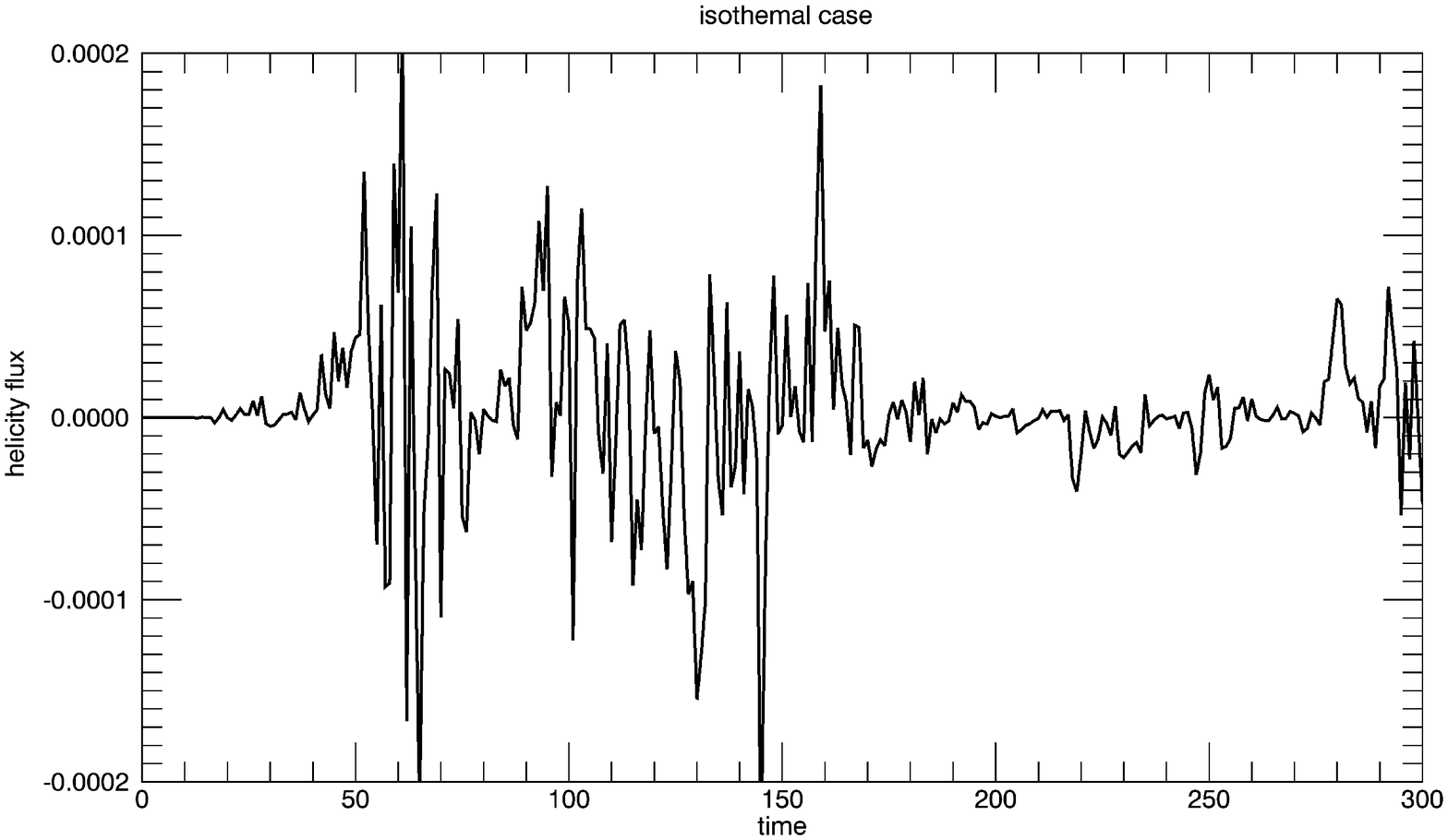} % requires the graphicx package
   \caption{Isothermal case. Time history of the helicity flux though the horizontal boundaries.}
   \label{fig:helflux_iso}
\end{figure}

We now turn to the isothermal case. Figure~\ref{fig:tothel_iso} and \ref{fig:helflux_iso} show respectively the total helicity and helicity flux as functions of time. A few things are immediately obvious, the average twist for the isothermal case is smaller than for the perfect-gas case, also the characteristic time-scale for fluctuations in the fluxes are much shorter than the corresponding time scale for the total helicity. In fact a simple comparison of the values shows that the contributions to the total helicity from boundary fluxes are all but negligible. Whatever twist is present is generated internally. This result is not surprising if we   recall that in the isothermal case the average density stratification is close to a Gaussian with most of the matter concentrated in a narrow region about the mid-plane. This is the region in which most of the magnetic energy and most of the Maxwell stresses are concentrated. Accordingly, the contributions to the helicity integral come from the same region with hardly anything flowing in or out of the boundaries. Inspection of the pdf for the isothermal case (blue line in Figure~\ref{fig:pdf}) shows that in analogy with the perfect gas case the value of the total helicity is mostly controlled by the symmetry of the pdf.

\section{Discussion } 
We have initiated a numerical study of the relationship between large-scale dynamo action and magnetic helicity in domains with  open magnetic boundaries. The purpose of this study is to provide some quantitative evidence to support the general idea that fluxes of magnetic helicity are an important ingredient in the large-scale dynamo process. The problem of the lack of gauge invariance in the definition of helicity and helicity fluxes in open domains has been circumvented by choosing a particular gauge, the winding gauge that, at least in domains with slab geometry, lends itself to a natural physical interpretation in terms of average pairwise twist of field lines \citep{Prior14}.  We have applied these ideas to two related systems consisting of dynamos driven by MRI turbulence in shearing boxes with zero net flux. The difference between the two systems was in the equation of state, isothermal in one case, perfect-gas in the other. The reason to pick these two particular system was because despite their obvious similarities, they are basically the same system except for the  equations of state, their dynamo properties are radically different. In the isothermal case there is little or no evidence for large-scale dynamo action, in terms of generation of substantial flux, whereas the evidence  in the perfect-gas case is very convincing. 

We find that there is no evidence for symmetry breaking with the formation of a helical state. Rather the helicity fluctuates, changing sign on a time-scale much shorter than the corresponding timescale for sign changes in the magnetic flux. Most of the helicity density is concentrated in isolated flux bundles whose average twist is of order unity. The presence of nearly equal amounts of the right and left handed bundles makes the average twist rather modest. In the perfect-gas case the helicity flux and helicity production rate are comparable, in the isothermal case the helicity flux is all but negligible. These considerations lead us to conclude that we find no compelling evidence of a simple relationship between large-scale dynamo action, measured, say, by the production of toroidal flux, and the helicity or the helicity flux. Even when substantial toroidal flux is present the overall state remains largely untwisted, and the helicity fluxes modest.  We would argue that it would be very difficult to use the time traces in figures \ref{fig:tothel_conv} or \ref{fig:tothel_iso} to place any useful constrain  on the dynamo efficiency. We should note that this result does not preclude that some quantity related to the magnetic vector potential other than the helicity may be a better indicator of large-scale dynamo action. We leave these speculations for a later study.

We close by suggesting that one possible explanation for tour findings  is that the system under consideration belongs to the class of essentially nonlinear dynamos in the sense that the velocity field responsible for the dynamo action is entirely magnetically driven--indeed it is the result of the growth of the magneto-rotational-instability to finite amplitude. The velocity in the unmagnetized state is purely a Keplerian shear and as such incapable of sustaining a dynamo. It is only through the intervention of magnetic stresses that turbulence develops that can regenerate the magnetic field.
This should be contrasted with the more conventional type of large-scale dynamos in which the basic flow is helical from the start (i.e. it lacks reflectional symmetry) and is capable of dynamo action. In this case one would expect the emergence of a non-reflectionally symmetric magnetic field in which the magnetic helicity and its flux may play a more prominent role. It would be interesting to repeat the kind of analysis presented here on such a system.

\section{Acknowledgment}
This work was supported in part  by the National Science Foundation 
sponsored Center for Magnetic Self Organization at the University of Chicago. 
We acknowledge that the results in this paper have been achieved using the 
PRACE Research Infrastructure resource FERMI based in Italy at  the Cineca Supercomputing Center. 

%\bibpunct[; ] {(}{)}{,}{}{}{,}

\appendix
\section{Magnetic helicity calculation}
In order to compute magnetic helicity and its flux,   equations (\ref{eq:vec-winding}) and (\ref{eq:phi}) must be solved to determine the vector and scalar potentials in the winding gauge. 
If all quantities were periodic this task could most easily be achieved by Fourier transform methods. However the data as it stands is not periodic in the $z$ direction and it is only shear periodic in $x$. Furthermore, even if ${\bf B}$ were periodic, {\bf A}, need not be. We show here how the data can easily be rearranged so that everything is exactly periodic. 
For the $x$ direction the problem can easily be overcome by restricting our analysis to the times when the shear-periodic domain re-aligns and the  data becomes exactly periodic in $x$.
Furthermore, the magnetic field can be split into two parts, namely
\begin{equation}
{\bf B}=\overline{{\bf B}}(z)+ {\bf b},
\label{eq:split}
\end{equation}
where $\overline{{\bf B}}$ is the horizontally averaged magnetic field. The vector potential associated with ${\bf b}$ is itself periodic and poses no problems, the vector potential associated with the average can be calculated separately and is given by (\ref{eq:AxAy}).
In the $z$ direction we can exploit some useful symmetries implied by the boundary conditions in $z$.
We recall that these are
\begin{equation}
B_x = 0,  \qquad  B_y = 0, \qquad \frac{\partial B_z}{\partial z} = 0
\end{equation}
and
\begin{equation}
\frac{\partial v_x}{\partial z} = 0,  \qquad  \frac{\partial v_y}{\partial z} = 0, \qquad B_z= 0
\end{equation}
on both boundaries. We can construct periodic data in the $z$ direction  by the following procedure. Given the original domain ${\cal D} = \{ 0  < x < L_x, 0 < y < L_y, 0 < z < L_z \}$, we define a new domain $\cal D'$, with double size in the vertical direction, ${\cal D'} =\{ 0 < x < L_x, 0 < y < L_y, 0 < z < 2 L_z \} $. For each function $f$ defined on $\cal D$, that satisfies the condition $f=0$ on the boundary, we define a function $f'$ on $\cal D'$ by copying the original function $f$ in the lower half of $\cal D'$, we then reverse $f$ along the $z$ direction, change sign and copy to the upper half of $\cal D'$. More precisely we define $f'$ as 
\begin{equation}
f' (x, y, z) = 
\begin {cases}
f(x, y, z) & \hbox{if}  \quad z < L_z \\
-f(x,y, 2 L_z - z) & \hbox{if} \quad L_z < z  < 2 L_z   
\end{cases}
\end{equation}
For each function $f$ defined on $\cal D$, that satisfies the condition $\partial f/\partial z=0$ on the boundary,we proceed in the same way, but we don't change sign. We have then
\begin{equation}
f' (x, y, z) = 
\begin {cases}
f(x, y, z) & \hbox{if}  \quad z < L_z \\
f(x,y, 2 L_z - z) & \hbox{if} \quad L_z < z  < 2 L_z .  
\end{cases}
\end{equation}
The functions defined on the domain  $\cal D'$ are  fully periodic and can be treated by fast Fourier transforms techniques.


\begin{thebibliography}{} 

\bibitem[\protect\citeauthoryear{{Balbus}}{{Balbus}}{2003}]{Balbus03}
{Balbus} S.~A.,  2003, \araa, 41, 555

\bibitem[\protect\citeauthoryear{{Berger} \& {Field}}{{Berger} \&
  {Field}}{1984}]{Berger84}
{Berger} M.~A.,  {Field} G.~B.,  1984, Journal of Fluid Mechanics, 147, 133

\bibitem[\protect\citeauthoryear{{Blackman} \& {Field}}{{Blackman} \&
  {Field}}{2000}]{Blackman00}
{Blackman} E.~G.,  {Field} G.~B.,  2000, \apj, 534, 984

\bibitem[\protect\citeauthoryear{{Blackman} \& {Field}}{{Blackman} \&
  {Field}}{2002}]{Blackman02}
{Blackman} E.~G.,  {Field} G.~B.,  2002, Physical Review Letters, 89, 265007

\bibitem[\protect\citeauthoryear{{Bodo}, {Cattaneo}, {Mignone} \&
  {Rossi}}{{Bodo} et~al.}{2012}]{Bodo12}
{Bodo} G.,  {Cattaneo} F.,  {Mignone} A.,    {Rossi} P.,  2012, \apj, 761, 116

\bibitem[\protect\citeauthoryear{{Bodo}, {Cattaneo}, {Mignone} \&
  {Rossi}}{{Bodo} et~al.}{2013}]{Bodo13}
{Bodo} G.,  {Cattaneo} F.,  {Mignone} A.,    {Rossi} P.,  2013, \apjl, 771, L23

\bibitem[\protect\citeauthoryear{{Bodo}, {Cattaneo}, {Mignone} \&
  {Rossi}}{{Bodo} et~al.}{2014}]{Bodo14}
{Bodo} G.,  {Cattaneo} F.,  {Mignone} A.,    {Rossi} P.,  2014, \apjl, 787, L13

\bibitem[\protect\citeauthoryear{{Bodo}, {Cattaneo}, {Mignone} \&
  {Rossi}}{{Bodo} et~al.}{2015}]{Bodo15}
{Bodo} G.,  {Cattaneo} F.,  {Mignone} A.,    {Rossi} P.,  2015, \apj, 799, 20

\bibitem[\protect\citeauthoryear{{Cattaneo} \& {Hughes}}{{Cattaneo} \&
  {Hughes}}{1996}]{Cattaneo96}
{Cattaneo} F.,  {Hughes} D.~W.,  1996, \pre, 54, R4532

\bibitem[\protect\citeauthoryear{{Davis}, {Stone} \& {Pessah}}{{Davis}
  et~al.}{2010}]{Davis10}
{Davis} S.~W.,  {Stone} J.~M.,    {Pessah} M.~E.,  2010, \apj, 713, 52

\bibitem[\protect\citeauthoryear{{Ebrahimi} \& {Bhattacharjee}}{{Ebrahimi} \&
  {Bhattacharjee}}{2014}]{Ebrahimi14}
{Ebrahimi} F.,  {Bhattacharjee} A.,  2014, Physical Review Letters, 112, 125003

\bibitem[\protect\citeauthoryear{{Field} \& {Blackman}}{{Field} \&
  {Blackman}}{2002}]{Field02}
{Field} G.~B.,  {Blackman} E.~G.,  2002, \apj, 572, 685

\bibitem[\protect\citeauthoryear{{Flaig}, {Kley} \& {Kissmann}}{{Flaig}
  et~al.}{2010}]{Flaig10}
{Flaig} M.,  {Kley} W.,    {Kissmann} R.,  2010, \mnras, 409, 1297

\bibitem[\protect\citeauthoryear{{Gressel}}{{Gressel}}{2010}]{Gressel10}
{Gressel} O.,  2010, \mnras, 405, 41

\bibitem[\protect\citeauthoryear{{Gressel}}{{Gressel}}{2013}]{Gressel13}
{Gressel} O.,  2013, \apj, 770, 100

\bibitem[\protect\citeauthoryear{{Gruzinov} \& {Diamond}}{{Gruzinov} \&
  {Diamond}}{1994}]{Diamond94}
{Gruzinov} A.~V.,  {Diamond} P.~H.,  1994, Physical Review Letters, 72, 1651

\bibitem[\protect\citeauthoryear{{Guan} \& {Gammie}}{{Guan} \&
  {Gammie}}{2011}]{Guan11}
{Guan} X.,  {Gammie} C.~F.,  2011, \apj, 728, 130

\bibitem[\protect\citeauthoryear{{Hirose}}{{Hirose}}{2015}]{Hirose15}
{Hirose} S.,  2015, \mnras, 448, 3105

\bibitem[\protect\citeauthoryear{{Hirose}, {Blaes}, {Krolik}, {Coleman} \&
  {Sano}}{{Hirose} et~al.}{2014}]{Hirose14}
{Hirose} S.,  {Blaes} O.,  {Krolik} J.~H.,  {Coleman} M.~S.~B.,    {Sano} T.,
  2014, \apj, 787, 1

\bibitem[\protect\citeauthoryear{{Kulsrud} \& {Anderson}}{{Kulsrud} \&
  {Anderson}}{1992}]{Kulsrud92}
{Kulsrud} R.~M.,  {Anderson} S.~W.,  1992, \apj, 396, 606

\bibitem[\protect\citeauthoryear{{Mignone}, {Bodo}, {Massaglia}, {Matsakos},
  {Tesileanu}, {Zanni} \& {Ferrari}}{{Mignone} et~al.}{2007}]{PLUTO}
{Mignone} A.,  {Bodo} G.,  {Massaglia} S.,  {Matsakos} T.,  {Tesileanu} O.,
  {Zanni} C.,    {Ferrari} A.,  2007, \apjs, 170, 228

\bibitem[\protect\citeauthoryear{{Moffatt}}{{Moffatt}}{1978}]{moffatbook}
{Moffatt} H.~K.,  1978, {Magnetic field generation in electrically conducting
  fluids}.
Cambridge, England, Cambridge University Press, 1978.~353 p.

\bibitem[\protect\citeauthoryear{{Prior} \& {Yeates}}{{Prior} \&
  {Yeates}}{2014}]{Prior14}
{Prior} C.,  {Yeates} A.~R.,  2014, \apj, 787, 100

\bibitem[\protect\citeauthoryear{{Shi}, {Krolik} \& {Hirose}}{{Shi}
  et~al.}{2010}]{Shi10}
{Shi} J.,  {Krolik} J.~H.,    {Hirose} S.,  2010, \apj, 708, 1716

\bibitem[\protect\citeauthoryear{{Shukurov}, {Sokoloff}, {Subramanian} \&
  {Brandenburg}}{{Shukurov} et~al.}{2006}]{Shukurov06}
{Shukurov} A.,  {Sokoloff} D.,  {Subramanian} K.,    {Brandenburg} A.,  2006,
  \aap, 448, L33

\bibitem[\protect\citeauthoryear{{Subramanian} \& {Brandenburg}}{{Subramanian}
  \& {Brandenburg}}{2004}]{Subramanian04}
{Subramanian} K.,  {Brandenburg} A.,  2004, Physical Review Letters, 93, 205001

\bibitem[\protect\citeauthoryear{{Sur}, {Shukurov} \& {Subramanian}}{{Sur}
  et~al.}{2007}]{Sur07}
{Sur} S.,  {Shukurov} A.,    {Subramanian} K.,  2007, \mnras, 377, 874

\bibitem[\protect\citeauthoryear{{Vainshtein} \& {Cattaneo}}{{Vainshtein} \&
  {Cattaneo}}{1992}]{Cattaneo92}
{Vainshtein} S.~I.,  {Cattaneo} F.,  1992, \apj, 393, 165

\bibitem[\protect\citeauthoryear{Vishniac \& Cho}{Vishniac \& Cho}{2001}]{VC01}
Vishniac E.~T.,  Cho J.,  2001, \apj, 550, 752

\end{thebibliography}
\end{document}